\documentclass[twoside,twocolumn,english,aps,prb,longbibliography]{revtex4-2}
\usepackage[T1]{fontenc}

\usepackage{amsmath}
\usepackage{amssymb}
\usepackage[english]{babel}
\makeatletter
\usepackage{tikz}
\usetikzlibrary{decorations.markings, arrows.meta, calc, backgrounds}

\makeatother

\usepackage[english]{babel}
\begin{document}
\title{Quantum Decoherence of the Surface Code: A Generalized Caldeira-Leggett
Approach}
\begin{abstract}
Standard quantum error correction (QEC) models typically assume discrete,
Markovian noise, obscuring the continuous quantum nature of physical
environments. In this manuscript, we investigate the fundamental limits
of an actively corrected surface code coupled to a continuous, un-reset
quantum environment at zero and finite temperature. Using the generalized
Caldeira-Leggett framework, we map the long-time evolution of the
logical qubit to a boundary conformal field theory, establishing an
exact equivalence to the anisotropic Kondo model. We evaluate computational
times for a finite code distance $L$ for all spatial and temporal
correlations. Our analysis reveals that a true thermodynamic threshold
exists strictly for short-range environments ($z>1/(s+1)$). In critical
or long-range regimes, the macroscopic footprint of the code weaponizes
the continuous bath, hindering the topological protection.
\end{abstract}
\author{E. Novais}
\email{corresponding author: eduardo.novais@ufabc.edu.br}
\affiliation{Centro de Ci\^{e}ncias Naturais e Humanas, Federal University of ABC, Brazil}

\author{A. H. Castro-Neto}
\affiliation{National University of Singapore, Institute for Functional Intelligent Materials, NUS S9 Building, 4 Science Drive 2, 117544 Singapore}
\affiliation{National University of Singapore, Centre for Advanced 2D Materials, 6 Science Drive 2, 117546 Singapore}
\affiliation{National University of Singapore, Department of Materials Science Engineering, 9 Engineering Drive 1, 117575 Singapore}
\affiliation{National University of Singapore, Department of Physics, 2 Science Drive 3, 117551 Singapore}

\date{\today{}}
\maketitle

\section{Introduction}

Quantum computation\citep{feynman_feynman_2001} holds the promise
to solve certain computational classes fundamentally faster than classical
systems\citep{nielsen_quantum_2012}. However, realizing this potential
requires protecting fragile quantum properties from a universe that
aggressively drives systems toward classicality. This quantum-classical
transition stalled the field for decades, and quantum computing only
became a mainstream proposition following the development of the theory
of quantum error correction (QEC)\citep{calderbank_good_1996,steane_simple_1996,terhal_quantum_2015,nielsen_quantum_2012}.
The key insight of QEC is to encode information within a highly entangled
logical Hilbert space defined across many physical qubits. Because
this logical space is spatially delocalized, it protects the quantum
information from the most common sources of local decoherence.

Among QEC protocols, the surface code\citep{dennis_topological_2002,wang_threshold_2010,fowler_surface_2012,terhal_quantum_2015,google_quantum_ai_suppressing_2023,vezvaee_surface_2025}
stands out as the preeminent candidate for fault-tolerant architecture.
By encoding logical qubits within the global topological properties
of a two-dimensional lattice, it physically manifests this spatial
delocalization, requiring a macroscopic string of localized errors
to span the lattice before the quantum information is compromised.

In the vast majority of QEC and fault-tolerance literature, this decoherence
is modeled purely as a stochastic, Markovian process. The environment
is treated essentially as a classical variable that applies random
Pauli errors to the physical qubits with a given probability. Under
this classical assumption, threshold theorems rigorously guarantee
that if the error rate remains sufficiently low, classical syndrome
extraction and active recovery operations can preserve the quantum
state indefinitely\citep{aliferis_quantum_2006,aharonov_fault-tolerant_2008}.
Furthermore, even advanced threshold theorems designed for coherent
noise rely on bounded operator norms to guarantee fault tolerance\citep{aharonov_fault-tolerant_2006},
a mathematical assumption that strictly fails, for instance, when
the logical qubit is coupled to the unbounded operators of a continuous
bosonic field theory. The stochastic approximation obscures a fundamental
physical reality: qubits do not interact with classical noise, but
rather with a continuous, interacting quantum environment. 

To capture this quantum reality, we must look beyond classical threshold
theorems to the foundational models of quantum dissipation\citep{weiss_quantum_2012,breuer_theory_2010}
and study adversarial quantum environments\citep{alicki_quantum_2004,dyakonov_is_2006,novais_resilient_2007,kalai_quantum_2009,kalai_how_2011,preskill_sufficient_2013,kalai_gaussian_2014}.
More than forty years ago, Caldeira and Leggett addressed a conceptually
identical problem\citep{caldeira_quantum_1983,caldeira_path_1983,leggett_quantum_1984,leggett_dynamics_1987,caldeira_dissipative_1993}.
Building upon the path-integral formalism of Feynman and Vernon\citep{feynman_theory_1963},
they modeled the environment as a continuous bath of quantum harmonic
oscillators. This framework profoundly advanced our understanding
of the quantum-classical transition by demonstrating exactly how macroscopic
variables can experience quantum tunneling. Because the logical qubit
of a surface code is inherently a macroscopic quantum variable, the
Caldeira-Leggett model provides the exact, generalized framework required
to understand its dissipative dynamics.

In this manuscript, we investigate the fundamental decoherence limits
of an actively corrected surface code when coupled to a continuous,
un-reset quantum environment. By integrating out the fast intra-cycle
dynamics, we map the long-time evolution of the logical qubit to a
one-dimensional boundary conformal field theory. Remarkably, we find
that the resulting effective Hamiltonian maps exactly to a macroscopic
spin-boson impurity model\citep{leggett_dynamics_1987} and not to
a set of competing environments\citep{castro_neto_quantum_2003,novais_frustration_2005}.
Our analysis reveals that while active quantum error correction successfully
filters out high-frequency ultraviolet fluctuations, the macroscopic
memory remains inherently vulnerable to the slow infrared modes of
the bath. We establish that a true thermodynamic threshold only exists
for environments with short-range spatial correlations, where the
dynamical exponent is $z>1/2$. Similar results were derived in a
different context before\citep{novais_resilient_2007,novais_hamiltonian_2008,novais_bound_2010,preskill_sufficient_2013,lopez-delgado_long-time_2017}.
Our approach is set apart from these previous results by considering
all possible logical errors, not resetting the environment at the
end of each QEC cycle, and writing computational time expressions
for all possible regimes. In critical ($z=1/2$) and long-range ($z<1/2$)
regimes, the system exhibits no strict asymptotic threshold; instead,
the renormalization group flow inevitably drives the loss of coherence.
Furthermore, we demonstrate that at finite temperatures, the ongoing
execution of the error correction cycle plausibly heats the low-frequency
environment, replacing idealized zero-temperature algebraic decay
with strictly exponential Korringa-like thermal relaxation, thereby
imposing a finite operational lifetime on the quantum memory across
all regimes for finite $L$.

The remainder of this paper is organized as follows. In Section II,
we review the theoretical construction of the surface code, detailing
the transition from a passive topological memory to an active quantum
error correction protocol. In Section III, we introduce our continuous
error model, utilizing the Dyson series to evaluate the intra-cycle
quantum evolution and formally deriving the macroscopic boundary interaction
between the logical qubit and the generalized Ohmic environment. Section
IV maps this resulting infrared theory to the anisotropic Kondo problem,
employing renormalization group (RG) equations to determine the zero-temperature
operational lifetimes across distinct spatial scaling regimes (short-range,
critical, and long-range) and discussing the implications for specific
hardware architectures, including superconducting circuits and neutral
atom arrays. In Section V, we extend this framework to finite temperatures,
demonstrating how the continuous execution of the QEC cycle introduces
a thermal cutoff that replaces algebraic decay with exponential Korringa-like
relaxation. Finally, Section VI provides our summary and conclusions,
while Appendix A presents a rigorous microscopic derivation of the
effective noise model from an adversarial itinerant fermionic bath.

\section{The Surface Code}

\begin{figure}[h]
	\centering
	\begin{tikzpicture}[scale=1.2]
		
		\fill[gray!40] (3,1) rectangle (4,2);
		
		\fill[gray!40] (1.5,2) -- (2,2.5) -- (2.5,2) -- (2,1.5) -- cycle;
		
		\foreach \x in {0, 1, 2, 3, 4, 5} {
			\draw[thick] (\x,0) -- (\x,3);
		}
		
		\foreach \y in {1, 2} {
			\draw[thick] (0,\y) -- (5,\y);
		}
		
		\foreach \x in {0.5, 1.5, 2.5, 3.5, 4.5} {
			\foreach \y in {1, 2} {
				\node at (\x,\y) {$\times$};
			}
		}
		
		\foreach \x in {0, 1, 2, 3, 4, 5} {
			\foreach \y in {0.5, 1.5, 2.5} {
				\node at (\x,\y) {$\times$};
			}
		}
		
		\node[fill=gray!40, inner sep=1.5pt] at (2,2) {$A_s$};
		\node[fill=gray!40, inner sep=1.5pt] at (3.5,1.5) {$B_p$};
		
	\end{tikzpicture}
	\caption{The surface code lattice. Physical qubits are located on the edges of the square lattice (crosses). The star operator $A_s$ acts on the four qubits sharing a vertex, while the plaquette operator $B_p$ acts on the four qubits enclosing a tile. The open boundaries distinguish the logical encoding.}
	\label{surface_code}
\end{figure}
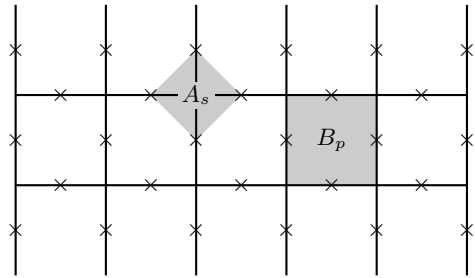

The surface code is arguably the simplest topological quantum error
correction protocol\citep{freedman_topological_2002,dennis_topological_2002}.
It maps exactly to a self-dual $\mathbb{Z}_{2}$ lattice gauge theory\citep{kogut_introduction_1979},
where physical qubits are located on the links of a two-dimensional
square lattice with open boundary conditions. Two basic gauge-invariant
operators define the model (see Fig.~\ref{surface_code}). Star operators
are the product of the four Pauli $\sigma_{i}^{x}$ matrices sharing
a vertex $s$ of the lattice,

\begin{equation}
\ensuremath{A_{s}=\prod_{i\in s}\sigma_{i}^{x}},
\end{equation}
while plaquette operators are the product of the four Pauli $\sigma_{i}^{z}$
matrices enclosing a tile $p$ of the lattice,
\begin{equation}
B_{p}=\prod_{i\in p}\sigma_{i}^{z}.
\end{equation}
At the boundaries of the lattice, these stabilizers reduce to three-qubit
operators. Because these boundary terms do not fundamentally alter
the bulk dissipative dynamics, we omit them from our present analysis\citep{novais_surface_2013}.

\begin{figure}[tbh]
	\centering
	\begin{tikzpicture}[scale=1.2]
		
		\draw[line width=12pt, gray!30] (-0.2, 1.5) -- (1.5, 1.5) -- (1.5, 2.5) -- (5.2, 2.5);
		\node at (-0.6, 1.5) {$\overline{X}$};
		
		\draw[line width=12pt, gray!50] (3, 3.2) -- (3, 2) -- (2, 2) -- (2, -0.2);
		\node at (2, -0.6) {$\overline{Z}$};
		
		\foreach \x in {0, 1, 2, 3, 4, 5} {
			\draw[thick] (\x,0) -- (\x,3);
		}
		
		\foreach \y in {1, 2} {
			\draw[thick] (0,\y) -- (5,\y);
		}
		
		\foreach \x in {0.5, 1.5, 2.5, 3.5, 4.5} {
			\foreach \y in {1, 2} {
				\node at (\x,\y) {$\times$};
			}
		}
		
		\foreach \x in {0, 1, 2, 3, 4, 5} {
			\foreach \y in {0.5, 1.5, 2.5} {
				\node at (\x,\y) {$\times$};
			}
		}
		
	\end{tikzpicture}
	\caption{Logical operators of the surface code. The macroscopic strings $\gamma_x$ and $\gamma_z$ span opposite boundaries of the lattice. The shaded paths indicate the physical qubits acted upon by the logical $\overline{X}$ and $\overline{Z}$ operators, representing non-trivial Wilson loops in the $\mathbb{Z}_2$ lattice gauge theory.}
	\label{logical_operators}
\end{figure}
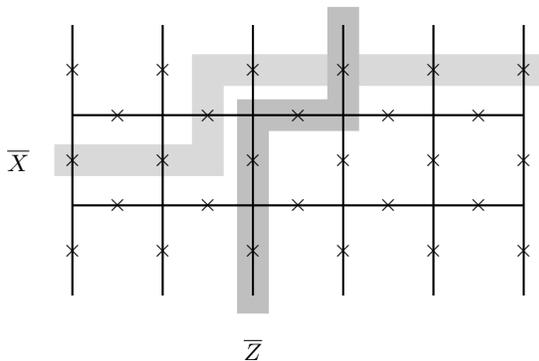

In the language of the $\mathbb{Z}_{2}$ lattice theory, the star
and plaquette operators act as the generators of trivial Wilson loops.
In addition to these local stabilizers, the code defines two logical
Pauli operators that correspond to non-trivial Wilson loops,
\begin{equation}
\overline{X}=\prod_{i\in\gamma_{x}}\sigma_{i}^{x},\quad\quad\overline{Z}=\prod_{i\in\gamma_{z}}\sigma_{i}^{z},
\end{equation}
where $\gamma_{x}$ and $\gamma_{z}$ are macroscopic continuous strings
of physical qubits that span opposite boundaries of the lattice (see
Fig.~\ref{logical_operators}).

This topological construction can be implemented using two distinct
QEC approaches: passive and active. Originally, the surface code was
conceived as a passive quantum memory. In this approach, the logical
Hilbert space is defined as the degenerate ground state of a local
Hamiltonian acting on an $L\times L$ lattice. The local Hamiltonian,

\begin{equation}
H_{\text{code}}=-\frac{\Delta_{0}}{2}\left(\sum_{s}A_{s}+\sum_{p}B_{p}\right),
\end{equation}
energetically enforces the $+1$ eigenvalue for all star and plaquette
stabilizers, and $2\Delta_{0}$ defines the macroscopic topological
gap protecting the code subspace. The two logical codewords are constructed
by projecting the fully polarized ferromagnetic state into the simultaneous
$+1$ eigenspace of all stabilizers

\begin{align}
\left|\bar{\uparrow}\right\rangle  & ={\cal N}\prod_{s}\left(1+A_{s}\right)\left|\uparrow\uparrow\uparrow...\uparrow\uparrow\uparrow\right\rangle ,\\
\left|\bar{\downarrow}\right\rangle  & =\bar{X}\left|\bar{\uparrow}\right\rangle ,
\end{align}
where $\mathcal{N}$ is a normalization constant.

This topological gap protects the encoded information from local physical
errors, modeled standardly as the independent stochastic application
of $\sigma_{i}^{x}$ or $\sigma_{i}^{z}$ with probability $p$. At
strictly zero temperature, the system remains confined to the ground
state manifold. Consequently, a logical bit or phase flip can only
occur via a macroscopic string of localized errors spanning the entire
lattice, an event whose probability $p^{L}$ is exponentially suppressed
by the code distance $L$. 

However, it is now well established that this two-dimensional topological
model is inherently unstable against finite thermal fluctuations,
a phenomenon known as thermal fragility\citep{alicki_thermalization_2009,bravyi_no-go_2009,terhal_quantum_2015}.
While creating an initial pair of anyonic excitations (the endpoints
of an open string) requires an energy penalty $2\Delta_{0}$, the
subsequent stretching of this string incurs no additional energy cost.
Because the model lacks string tension, thermal fluctuations effortlessly
drive a random walk of anyons to the lattice edges, destroying the
stored quantum information when a logical operator is created.

To overcome this fundamental thermodynamic limitation, modern architectures
employ an active QEC protocol. Rather than relying on a passive energy
gap to confine anyons, the system undergoes discrete projective measurements
of the stabilizers $A_{s}$ and $B_{p}$ at regular intervals, defining
the syndrome extraction cycle $\tau_{\text{QEC}}$. In this approach,
there is no physical $H_{\text{code}}$ acting continuously; instead,
the system is periodically measured and projected back into the logical
Hilbert space. A classical decoding algorithm then pairs the resulting
anyons and determines the optimal recovery operation, effectively
neutralizing the thermal random walk.

By executing this active cycle much faster than the typical timescale
of thermal diffusion, active QEC successfully circumvents thermal
fragility. Provided the physical error rate remains below a specific
threshold, this active intervention theoretically guarantees an infinitely
long-lived quantum memory\citep{fowler_surface_2012}. However, this
standard active threshold framework relies on a crucial approximation:
it models decoherence strictly as classical, stochastic Markovian
noise.

The assumption of a classical environment obscures the true quantum
dynamics experienced by the code. The syndrome extraction plays a
critical thermodynamic role: by projectively collapsing the local
quantum state, the measurement reduces the entropy of the system,
fundamentally excluding quantum interference between dynamic error
paths associated with incompatible error syndromes\citep{novais_decoherence_2006,novais_resilient_2007}.
This projective collapse introduces a fundamental scale separation
between the intra-cycle and inter-cycle quantum dynamics\citep{novais_resilient_2007,novais_decoherence_2006,novais_hamiltonian_2008,novais_bound_2010}.

Previously, we analyzed this inter-cycle quantum dynamics under the
explicit assumption that the environment is reset at each syndrome
extraction\citep{novais_resilient_2007,novais_decoherence_2006,novais_hamiltonian_2008,novais_bound_2010,jouzdani_fidelity_2014}.
Those previous works focused primarily on the role of spatial correlations
mediated by the quantum bath and, consequently, relied on the assumption
of this environmental reset to truncate the temporal memory. Here,
we remove that restriction. We shift our focus to the long-time dynamics
of the logical Hilbert space and the continuous, unbroken loss of
quantum information to an un-reset quantum environment.

\section{The Error Model}

In an ideal active QEC architecture, the physical qubits possess no
intrinsic Hamiltonian dynamics; their evolution is dictated entirely
by the application of quantum gates and projective measurements. To
rigorously isolate the decoherence induced by the environment, we
assume that these QEC operations are structurally perfect and occur
instantaneously at discrete clock times $t=n\tau_{\text{QEC}}$, where
$n$ is an integer.

Consequently, the qubits evolve freely during the open intervals between
these discrete clock ticks. It is exclusively during these intra-cycle
periods of duration $\tau_{\text{QEC}}$ that the physical qubits
are exposed to the continuous dynamics of the quantum environment.
Because this model deliberately neglects the inevitable errors introduced
by finite gate times and imperfect classical measurements, the coherence
limits derived in this manuscript establish a strict upper bound on
the available quantum computation time\citep{novais_bound_2010}.

During the continuous intra-cycle time interval, the physical qubits
evolve unitarily under the influence of the environment. Because the
theoretical framework of QEC relies fundamentally on the assumption
that local physical errors act perturbatively, it is natural to describe
this quantum evolution using the Dyson series in the interaction picture.
The time-evolution operator governing a single QEC cycle is formally
given by

\begin{equation}
\hat{U}\left(\tau_{\text{QEC}},0\right)=T_{t}\exp\left[\frac{-i}{\hbar}\sum_{\vec{x}}\int_{0}^{\tau_{QEC}}dt\,\vec{f}\left(\vec{x},t\right)\cdot\vec{\sigma}_{\vec{x}}\right],
\end{equation}
where $T_{t}$ is the time-ordering operator and $\vec{x}$ denotes
the discrete spatial coordinates of the physical qubits on the lattice.
The vector $\vec{\sigma}_{\vec{x}}=(\sigma_{\vec{x}}^{x},\sigma_{\vec{x}}^{y},\sigma_{\vec{x}}^{z})$
contains the standard Pauli matrices acting on the qubit at position
$\vec{x}$, while the operator-valued field $\vec{f}(\vec{x},t)$
represents the local quantum coupling between the qubits and the environmental
degrees of freedom.

At the clock time $t=\tau_{\text{QEC}}$, the continuous unitary evolution
is abruptly halted by the simultaneous measurement of all star and
plaquette stabilizers. A classical decoding algorithm processes the
resulting syndrome to determine the optimal recovery operation. Physically,
this discrete intervention acts as a strict projective filter on the
Dyson series. The measurement collapses the quantum state, perfectly
annihilating all quantum amplitudes corresponding to dynamic error
trajectories that are incompatible with the extracted syndrome\citep{novais_decoherence_2006}.

It is critical to address the back-action of the classical syndrome
measurements on this continuous environment. The discrete QEC operations,
gates and projective measurements, are executed on an ultra-fast timescale,
$\tau_{\text{QEC}}$, which will serve as the strict ultraviolet (UV)
cutoff for our effective field theory. The measurement back-action
is therefore confined to the high-frequency, localized modes of the
bath. Consequently, the slow phase memory of the continuous environment
remains unbroken across discrete QEC cycles, rigorously validating
our assumption of an un-reset infrared bath.

In a purely passive memory, the lowest-order term in the Dyson series
capable of producing a logical error is of order $L$, the linear
dimension of the lattice. However, standard active decoding algorithms,
such as Minimum Weight Perfect Matching\citep{dennis_topological_2002},
inherently assumes that any extracted syndrome was generated by the
shortest possible physical error chain. This assumption introduces
a fundamental vulnerability due to the superposition of quantum histories
in the intra-cycle evolution.

\begin{figure}[h]
	\centering
	\begin{tikzpicture}[scale=0.8]
		
		\draw[line width=12pt, gray!30] (3, 0) -- (3, 1);
		\draw[line width=12pt, gray!30] (3, 4) -- (3, 6);
		
		\foreach \x in {0, 1, 2, 3, 4, 5, 6, 7} {
			\draw[thick] (\x,0) -- (\x,6);
		}
		
		\foreach \y in {1, 2, 3, 4, 5} {
			\draw[thick] (0,\y) -- (7,\y);
		}
		
		\foreach \x in {0.5, 1.5, 2.5, 3.5, 4.5, 5.5, 6.5} {
			\foreach \y in {1, 2, 3, 4, 5} {
				\node at (\x,\y) {$\times$};
			}
		}
		
		\foreach \x in {0, 1, 2, 3, 4, 5, 6, 7} {
			\foreach \y in {0.5, 1.5, 2.5, 3.5, 4.5, 5.5} {
				\node at (\x,\y) {$\times$};
			}
		}
		
		\fill[gray!60] (2.5,1) -- (3,1.5) -- (3.5,1) -- (3,0.5) -- cycle;
		\fill[gray!60] (2.5,4) -- (3,4.5) -- (3.5,4) -- (3,3.5) -- cycle;
		
		\node[fill=gray!60, inner sep=1.5pt] at (3,1) {$A_s$};
		\node[fill=gray!60, inner sep=1.5pt] at (3,4) {$A_{s^\prime}$};
		
	\end{tikzpicture}
	\caption{Second quantum amplitude generating the identical syndrome configuration. The bath introduces physical $Z$-errors that connect the anyons to the open boundaries. Because the classical decoder typically assumes the shortest path, an exact equivalence in length (as shown here at order $L/2=3$) creates strict ambiguity, inadvertently leading the decoder to complete a macroscopic logical $\overline{Z}$ operator.}
	\label{fig:amplitude_long}
\end{figure}
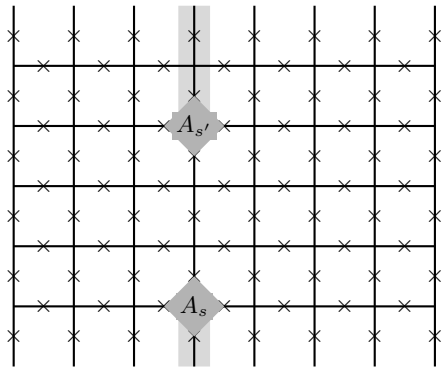

Suppose the extracted syndrome is consistent with a short physical
error chain of length $n<L/2$ (see Fig.~\ref{fig:amplitude_short}).
To neutralize the anyons, the classical decoder will apply a recovery
string of exactly length $n$. While this successfully corrects the
order-$n$ amplitude in the Dyson series by forming a trivial closed
loop, the un-projected quantum state also contains a complementary
amplitude. This complementary quantum path, which produces the identical
syndrome by connecting the anyons to the opposite boundaries, is of
order $L-n$ (see Fig.~\ref{fig:amplitude_long}).

\begin{figure}[h]
	\centering
	\begin{tikzpicture}[scale=0.8]
		
		\draw[line width=12pt, gray!30] (3, 1) -- (3, 4);
		
		\foreach \x in {0, 1, 2, 3, 4, 5, 6, 7} {
			\draw[thick] (\x,0) -- (\x,6);
		}
		
		\foreach \y in {1, 2, 3, 4, 5} {
			\draw[thick] (0,\y) -- (7,\y);
		}
		
		\foreach \x in {0.5, 1.5, 2.5, 3.5, 4.5, 5.5, 6.5} {
			\foreach \y in {1, 2, 3, 4, 5} {
				\node at (\x,\y) {$\times$};
			}
		}
		
		\foreach \x in {0, 1, 2, 3, 4, 5, 6, 7} {
			\foreach \y in {0.5, 1.5, 2.5, 3.5, 4.5, 5.5} {
				\node at (\x,\y) {$\times$};
			}
		}
		
		\fill[gray!60] (2.5,1) -- (3,1.5) -- (3.5,1) -- (3,0.5) -- cycle;
		\fill[gray!60] (2.5,4) -- (3,4.5) -- (3.5,4) -- (3,3.5) -- cycle;
		
		\node[fill=gray!60, inner sep=1.5pt] at (3,1) {$A_s$};
		\node[fill=gray!60, inner sep=1.5pt] at (3,4) {$A_{s^\prime}$};
		
	\end{tikzpicture}
	\caption{First quantum amplitude generating a specific syndrome configuration on an expanded lattice. The continuous quantum bath introduces a localized string of physical $Z$-errors (shaded path) that directly connects two star operators, resulting in two well-separated anyonic excitations (shaded diamonds).}
	\label{fig:amplitude_short}
\end{figure}
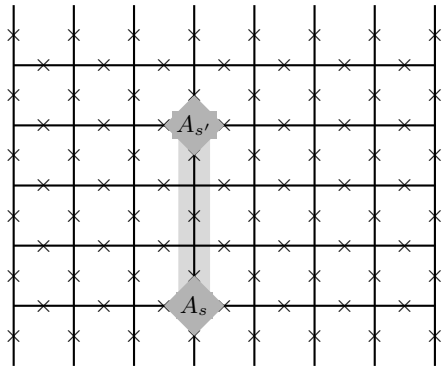

When the classical decoder applies its length-$n$ recovery operation,
it inadvertently combines with this order-$(L-n)$ amplitude to complete
a macroscopic, non-trivial Wilson loop across the lattice. Therefore,
resolving a syndrome with an assumed error of length $n$ inevitably
seals a logical error stemming from the quantum amplitude of order
$L-n$. The most critical situation for the system, the lowest-order
logical failure surviving in the Dyson series, occurs when this complementary
amplitude is minimized, which happens exactly at order $L/2$.

To proceed with the evaluation of the Dyson series, we must specify
the microscopic structure of the environmental coupling $\vec{f}(\vec{x},t)$.
Following the generalized Caldeira-Leggett framework, we model the
quantum bath as a continuous set of non-interacting harmonic oscillators
that couple linearly to the physical qubits. We begin our analysis
by considering a pure dephasing mechanism\citep{unruh_maintaining_1995,breuer_theory_2010},
where the environment interacts exclusively with the longitudinal
spin components; however, we will subsequently demonstrate how generalized
transverse errors dynamically emerge from this underlying interaction.
This initial dephasing model yields the standard interaction-picture
Hamiltonian

\begin{align}
\hat{H}_{\text{int}}(t) & =\sum_{\vec{x}}B(\vec{x},t)\sigma_{\vec{x}}^{z},\\
\hat{B}(\vec{x},t) & =\sum_{\vec{k}}\left(g_{\vec{k}}\,e^{i(\vec{k}\cdot\vec{x}-\omega_{\vec{k}}t)}a_{\vec{k}}+g_{\vec{k}}^{*}\,e^{-i(\vec{k}\cdot\vec{x}-\omega_{\vec{k}}t)}a_{\vec{k}}^{\dagger}\right)
\end{align}
where $a_{\vec{k}}^{\dagger}$ and $a_{\vec{k}}$ are the bosonic
creation and annihilation operators for the environmental mode with
momentum $\vec{k}$ and frequency $\omega_{\vec{k}}$, and $g_{\vec{k}}$
determines the momentum-dependent coupling strength.

Substituting the pure dephasing Hamiltonian into the Dyson series,
we expand the time-evolution operator to evaluate the discrete quantum
amplitudes. The $n$-th order term in this expansion involves $n$
time-ordered interactions between the lattice qubits and the continuous
bath. As established by the threshold ambiguity of the classical decoder,
the leading-order term capable of sealing a macroscopic logical error
occurs at $n=L/2$.

Extracting this specific order from the Dyson series yields
\begin{equation}
\hat{U}^{(L/2)}=\left(\frac{-i}{\hbar}\right)^{L/2}T_{t}\int_{0}^{\tau_{\text{QEC}}}\mathcal{D}t\left[\hat{H}_{\text{int}}(t_{1})\cdots\hat{H}_{\text{int}}(t_{L/2})\right],
\end{equation}
where $\int\mathcal{D}t$ denotes the nested time integration over
the cycle duration $\tau_{\text{QEC}}$ and $\hat{{\cal O}}$ are
operators in the interaction picture.

Inserting the microscopic definition of $\hat{H}_{\text{int}}(t)$,
we obtain a sum over all possible spatial configurations of $L/2$
local Pauli errors. However, the projective measurement of the active
QEC cycle annihilates the vast majority of these terms. The surviving
amplitudes are only those whose spatial coordinates $\{\vec{x}_{1},\vec{x}_{2},\dots,\vec{x}_{L/2}\}$
form a contiguous string $\gamma_{L/2}$ that matches the extracted
syndrome.

For a given critical path $\gamma_{L/2}$, the effective quantum amplitude
is given by the product of the physical qubit operators coupled to
a highly non-local, multi-point bath correlation function\begin{widetext}
\begin{equation}
\hat{U}_{\gamma}^{(L/2)}\propto\left(\prod_{j=1}^{L/2}\sigma_{\vec{x}_{j}}^{z}\right)T_{t}\int\mathcal{D}t\,\left[\hat{B}(\vec{x}_{1},t_{1})\hat{B}(\vec{x}_{2},t_{2})\cdots\hat{B}(\vec{x}_{L/2},t_{L/2})\right],
\end{equation}
\end{widetext}where $\int\mathcal{D}t$ denotes the time-ordered
integration over the cycle duration $\tau_{\text{QEC}}$.

This mathematical structure reveals the true thermodynamic threat
to the quantum memory. While the classical decoder trivially applies
the complementary string $\prod\sigma^{z}$ to complete the logical
$\overline{Z}$ operator, the quantum weight of this failure is entirely
dictated by the $L/2$-point temporal correlation function of the
bosonic bath. 

To extract the long-time asymptotic behavior of the quantum evolution,
we normal-order the time-ordered operator $\hat{U}_{\gamma}^{(L/2)}$\citep{polchinski_string_2005}.
In real space, this procedure is formally equivalent to the Operator
Product Expansion (OPE), which evaluates the temporal integral by
performing all possible Wick contractions of the $\hat{B}$ operators.
By assuming a specific two-point correlation function for the environment,
we can rigorously bound the magnitude of these contractions.

Because our primary interest lies in the long-time dynamics characteristic
of Ohmic dissipation, we model the environment as a continuous bath
in $D$ spatial dimensions. We define this bath with a dispersion
relation $\omega_{k}=v\left|k\right|^{z}$ and a momentum-dependent
coupling $g_{k}\propto k^{\alpha}$. In the generalized Caldeira-Leggett
framework, the environmental influence is fundamentally dictated by
the spectral density $J(\omega)\propto\sum_{\vec{k}}|g_{\vec{k}}|^{2}\delta(\omega-\omega_{\vec{k}})$.
To guarantee strictly Ohmic behavior, where $J(\omega)\propto\omega$,
these microscopic parameters must satisfy the structural constraint
$D+2\alpha=2z$. 

Imposing this constraint ensures that the general two-point correlation
functions follow the expected asymptotic forms,

\begin{align}
\left\langle T_{t}\hat{B}(\vec{x},t_{1})\hat{B}(\vec{x},t_{2})\right\rangle  & \propto\frac{\lambda^{2}}{v^{2}|t_{1}-t_{2}|^{2}},\label{eq:density-density}\\
\left\langle T_{t}\hat{B}(\vec{x}_{1},t)\hat{B}(\vec{x}_{2},t)\right\rangle  & \propto\frac{\lambda^{2}}{a_{0}^{2\left(1-z\right)}|\vec{x}_{1}-\vec{x}_{2}|^{2z}},
\end{align}
where $v$ is the bosonic velocity, $\lambda$ is an effective microscopic
coupling with dimensions of energy times length, and $a_{0}$ is a
characteristic short-distance cutoff of the bath.

Applying Wick's theorem to the $L/2$-point correlation function,
we decompose the time-ordered expectation value into a sum over all
fully contracted products of two-point propagators. For a string of
$L/2$ local interactions, there are exactly $(L/2-1)!!$ distinct
pairings. This factorial proliferation of Feynman diagrams represents
a massive combinatorial enhancement of the logical error amplitude.
However, this explosive growth is fundamentally opposed by the spatial
and temporal decay of the correlation functions. Each individual contraction
between two distinct points along the error string, $(\vec{x}_{i},t_{i})$
and $(\vec{x}_{j},t_{j})$, is penalized by the propagator.

By restricting the most relevant logical error to a single 1D contour
along the code geometry, we drastically simplify the spatial dependence
of these contractions. For a critical path $\gamma_{L/2}$ forming
a string, the coordinates of the physical qubits align such that the
distance between any two errors is simply $\left|\vec{x}_{i}-\vec{x}_{j}\right|=a\left|i-j\right|$,
where $a$ is the lattice constant. To evaluate the asymptotic scaling
of this logical error amplitude, we must substitute the generalized
Ohmic correlation functions into the Wick expansion. The temporal
integration over the discrete cycle duration $\tau_{\text{QEC}}$
yields an effective spatial pairwise contraction that scales with
the dynamical exponent $z$
\[
\mathcal{G}(x_{ij})\propto\frac{\lambda^{2}\tau_{\text{QEC}}^{2}}{a_{0}^{2(1-z)}a^{2z}\left|i-j\right|^{2z}}
\]
The total quantum amplitude for the macroscopic failure is determined
by the sum over all possible perfect matchings of these pairs along
the string. 

Crucially, this evaluated amplitude represents the quantum weight
of only a single, fully specified error configuration. To capture
the true macroscopic vulnerability of the quantum memory, we must
account for the massive degeneracy of failure modes introduced by
the classical decoder. Along any given one-dimensional contour of
length $L$, the continuous environment can distribute $L/2$ physical
errors in exactly $\binom{L}{L/2}$ distinct spatial configurations.
Because each of these configurations leaves exactly $L/2$ clean segments
interspersed along the string, the minimum-weight matching algorithm
inevitably seals the macroscopic logical error. Furthermore, the two-dimensional
lattice contains exactly $L$ independent, parallel contours where
this critical one-dimensional failure can take root. The total number
of pathways that produce the exact same logical error is therefore
enhanced to $L\binom{L}{L/2}$. Applying Stirling's approximation
in the macroscopic limit, we find that this combined combinatorial
and transverse entropy diverges as
\[
N_{\text{paths}}\approx\sqrt{\frac{2L}{\pi}}2^{L},
\]
representing a massive number of configurations to be considered.

For regimes where $2z>1$, the correlation strength decays rapidly,
and the dominant contribution arises strictly from pairing nearest-neighbor
errors. The sum converges and the effective weight of all Wick contractions
is just a finite number independent of $L$. However, this localized
approximation breaks down for smaller dynamical exponents where the
bath mediates long-range interactions across the lattice. Specifically,
when $z=1/2$, the spatial correlation decays as $1/x_{ij}$. Summing
the contributions of these long-range pairings across the one-dimensional
error string of length $L/2$ introduces a logarithmic divergence.
In this critical regime, the effective action of the logical error
path accumulates a scaling factor proportional to $\ln L$, making
the transition amplitude highly sensitive to the macroscopic dimensions
of the quantum memory. Thus, there are three distinct cases to consider

\[
\bar{\lambda}_{z}^{2}\propto\begin{cases}
\frac{16\lambda^{2}\tau_{\text{QEC}}^{2}}{\hbar^{2}a_{0}^{2(1-z)}a^{2z}} & z>1/2\\
\frac{16\lambda^{2}\tau_{\text{QEC}}^{2}}{\hbar^{2}a_{0}a}\ln L & z=1/2\\
\frac{16\lambda^{2}\tau_{\text{QEC}}^{2}}{\hbar^{2}a_{0}^{2(1-z)}a^{2z}}L^{1-2z} & z<1/2
\end{cases}
\]
After integrating all environmental modes in a QEC cycle, the operator
that couples the logical error with the infrared environment is 

\[
\hat{U}_{\gamma}^{(L/2)}\propto\frac{\lambda\tau_{\text{QEC}}}{\hbar}\sqrt{\frac{2L}{\pi}}\left(\bar{\lambda}_{z}^{2}\right)^{\frac{L}{4}}:\hat{{\cal B}}(t=0):
\]
where $:\dots:$ stands for the normal-ordering operator. To linear
order in the environmental coupling, the surviving slow modes are
now restricted to low-frequency modes, with $\omega_{q}$ running
from $0$ up to the new ultraviolet cutoff of the effective theory,
$\Lambda=1/\tau_{\text{QEC}}$. Since the spatial structure of the
code was completely integrated out, the logical qubit in the infrared
theory will experience Ohmic correlations with itself at later QEC
cycles. Formally, this infrared limit reveals that the long-time dynamics
of the logical qubit are strictly equivalent to a one-dimensional
continuous bosonic field theory interacting with a localized macroscopic
operator $\overline{Z}$.

Following standard 1D bosonization conventions\citep{delft_bosonization_1998},
we express this low-energy bath operator as the spatial gradient of
a continuous phase field evaluated at the origin

\begin{align}
\hat{{\cal B}}\left(t\right) & =\partial_{x}\hat{\phi}\left(0,t\right),
\end{align}
To recover the Ohmic spectral density, the bosonic field $\phi\left(x\right)$
is explicitly defined as
\begin{equation}
\phi\left(x\right)=\frac{1}{\sqrt{D}}\sum_{q\neq0}\frac{1}{\sqrt{\left|q\right|}}\left(e^{iqx}a_{q}+e^{-iqx}a_{q}^{\dagger}\right),
\end{equation}
where $\left|q\right|=\omega_{q}/v$ and $D$ serves as the macroscopic
infrared spatial regulator of the environment.

The dynamics of this continuous free bosonic bath are governed by
the standard Tomonaga-Luttinger Hamiltonian

\begin{equation}
H_{0}=\frac{v}{2}\int_{-\infty}^{\infty}dx\left\{ \left[\partial_{x}\phi\left(x\right)\right]^{2}+\left[\partial_{x}\theta\left(x\right)\right]^{2}\right\} ,\label{eq:H0}
\end{equation}
where $\theta(x)$ is the dual conjugate field, satisfying the fundamental
commutation relation where $\left[\phi\left(x\right),\partial\theta\left(y\right)\right]=i\delta\left(x-y\right)$.
Even though we assume the traditional Tomonaga-Luttinger Hamiltonian,
it is possible to generalize this environment by adding an even larger
dissipative environment that interacts with the free bosonic theory
of Eq.~(\ref{eq:H0})\citep{castro_neto_open_1997}. 

By integrating out the fast intra-cycle modes and absorbing the combinatorial
complexity of the classical decoder into the effective coupling, the
resulting infrared theory for the logical qubit decoherence reduces
to a remarkably compact boundary interaction model

\begin{equation}
H_{\text{IR}}=H_{0}+J_{z}\partial_{x}\phi\left(0\right)\bar{Z},
\end{equation}
Here, $J_{z}$ is the renormalized infrared coupling constant that
encapsulates the entire macroscopic scaling of the active QEC protocol,

\begin{equation}
J_{z}=\lambda\sqrt{\frac{2L}{\pi}}\left(\bar{\lambda}_{z}^{2}\right)^{\frac{L}{4}}.\label{eq:J_z}
\end{equation}
As is standard in boundary conformal field theory, this non-chiral
bath can be decomposed into symmetric and antisymmetric continuous
fields. Because the logical error can be interpreted as an impurity
scattering center (see Appendix~\ref{sec:An-adversary-noise} for
an explicit derivation), the antisymmetric channel completely decouples
from it and can be safely traced out, leaving the dissipative dynamics
governed entirely by the symmetric phase field. However, for mathematical
convenience in the subsequent notation, we do not explicitly trace
out the antisymmetric field, as its inclusion as a free background
mode does not alter the boundary dynamics or the instanton solutions.

To understand the profound physical implications of this boundary
interaction, we examine the static solutions to the classical equation
of motion. Using the standard mapping to the Lagrangian formalism,
we find

\begin{equation}
\partial_{t}\phi(x,t)=v\partial_{x}\theta\left(x,t\right),
\end{equation}
that gives the Lagrangian density, 
\begin{equation}
\mathcal{L}=\frac{1}{2v}(\partial_{t}\phi)^{2}-\frac{v}{2}(\partial_{x}\phi)^{2}-J_{z}\partial_{x}\phi\,\delta(x)\overline{Z}.
\end{equation}
In the low-energy limit, the temporal derivatives vanish, reducing
the dynamics to a purely spatial constraint
\begin{equation}
v\partial_{x}^{2}\phi(x)=-J_{z}\overline{Z}\partial_{x}\delta(x).
\end{equation}
Integrating this equation over the spatial domain reveals that the
logical operator rigidly dictates the macroscopic configuration of
the environmental field. Assuming antisymmetric boundary conditions
at infinity, the static solution manifests as a discontinuous step
profile centered at the origin
\[
\phi_{\text{vac}}(x)=-\frac{J_{z}}{2v}\overline{Z}\text{sgn}(x).
\]
Because the logical operator possesses two eigenvalues, $\overline{Z}=\pm1$,
this exact solution demonstrates that the interacting system possesses
two distinct, degenerate vacuum states. The continuous environment
dynamically reorganizes into one of two topological field configurations,
completely slaved to the logical state $|0\rangle_{L}$ or $|1\rangle_{L}$.
Crucially, these two topological vacua are separated by a macroscopic
field deformation. A transition between the logical states requires
the environment to globally invert its static phase field from $+\text{sgn}(x)$
to $-\text{sgn}(x)$. Such a massive collective reorganization cannot
occur via local, perturbative bosonic fluctuations. Instead, the transition
must proceed non-perturbatively through macroscopic quantum tunneling.
In the path-integral formulation, this tunneling between degenerate
vacua is governed by instantons, localized kink solutions in imaginary
time\citep{caldeira_influence_1981,castro_neto_alternative_1992,castro_neto_transport_1993}. 

To mathematically formalize these tunneling events as quantum operators
acting on the logical subspace, we must construct the physical transverse
error. The necessity of defining the physical transverse error via
the dual field $\theta$ is strictly dictated by the underlying algebraic
duality of the system. In the bare logical subspace, the bit-flip
and phase-flip operators are dual observables that must satisfy the
fundamental anticommutation relation, $\{\overline{X},\overline{Z}\}=0$.
However, in the infrared limit of the continuous error correction,
the logical $\overline{Z}$ operator does not act in isolation; it
is dynamically bound to the local gradient of the phase field, $\partial_{x}\phi(0)$.

To represent a true physical transition, the fully dressed dual error
must not only flip the bare logical spin but also execute the complementary
topological operation on the continuous bath. This is where the canonical
commutation relation of the 1D bosonic theory, $[\phi(x),\partial_{y}\theta(y)]=i\delta(x-y)$,
enforces the physics. Because $\theta$ is the conjugate to $\phi$,
the vertex operator $\exp(i\beta\hat{\theta}(0))$ acts as the exact
quantum mechanical generator of translation for the phase field.

Therefore, constructing the physical transverse error as $\overline{X}_{\text{phys}}\propto\overline{X}\cos(\sqrt{4\pi}\theta(0))$
uniquely satisfies both the discrete symmetries of the logical subspace
and the continuous symmetries of the bosonic environment. It guarantees
that as the bare $\overline{X}$ flips the macroscopic spin, the conjugate
vertex operator simultaneously translates the surrounding continuous
field between its two degenerate topological vacua, perfectly preserving
the global duality of the coupled quantum system.

The most general infrared model that is consistent with this physics
is 

\begin{widetext}
\begin{align}
H_{\text{IR}} & =H_{0}+J_{x}\cos\left(\sqrt{4\pi}\theta(0)\right)\bar{X}+J_{y}\sin\left(\sqrt{4\pi}\theta(0)\right)\bar{Y}+J_{z}\partial_{x}\phi\left(0\right)\bar{Z}.\label{eq:logical_kondo-1}
\end{align}
\end{widetext}Equation (\ref{eq:logical_kondo-1}) is precisely the
Kondo model\citep{hewson_kondo_2009} for an impurity spin-1/2 (the
logical qubit). 

It is important to emphasize that the precise microscopic origin of
this transverse coupling is largely immaterial to the macroscopic
dynamics due to the universality of the infrared (IR) limit. In any
realistic physical architecture, the macroscopic qubit is subject
to a multitude of microscopic noise channels, including unavoidably
non-commuting errors such as stray transverse fields or amplitude
damping. While these errors possess distinct microscopic UV signatures,
the renormalization group flow ensures that their low-energy physics
converges to a single universality class. Because any physical operator
that induces a logical bit-flip ($\bar{X}$) must preserve the canonical
commutation relations with the macroscopic phase field $\phi(x)$,
it must necessarily couple to the conjugate momentum field $\theta(x)$.
Consequently, regardless of the specific microscopic term that seeds
the macroscopic quantum tunneling, the effective IR dynamics of any
transverse bath noise are universally governed by the vertex operator
$\cos(\sqrt{4\pi}\theta(0))$.

The microscopic analysis in Appendix~A is an example of such an 
adversarial environment. Regardless of its specific fermionic 
or bosonic nature, it necessarily entangles with the logical 
subspace to produce the universal infrared limit of 
Eq.~(\ref{eq:logical_kondo-1}). However, the first-principles 
derivation presented here is fundamentally more profound: 
it explicitly demonstrates how the macroscopic entanglement 
between the topological Hilbert space and a continuous 
quantum environment dynamically emerges from the underlying
symmetries.

\section{The Kondo Problem for the Logical Qubit }

In the previous section, we established that the infrared model governing
the time evolution of a logical qubit in the presence of a continuous
quantum generalized Ohmic environment is formally identical to the
anisotropic Kondo Hamiltonian\citep{leggett_dynamics_1987}.

A particularly robust approach to evaluating the low-energy dynamics
of this model is to construct the partition function at zero temperature
in imaginary time, and subsequently Wick rotate the results back to
real time. Following this perturbative scaling approach\citep{anderson_exact_1969,affleck_kondo_1991,hewson_kondo_2009},
we arrive at the standard one-loop Renormalization Group (RG) equations

\begin{align}
\frac{\partial j_{x}}{\partial l} & =j_{y}j_{z},\\
\frac{\partial j_{y}}{\partial l} & =j_{x}j_{z},\\
\frac{\partial j_{z}}{\partial l} & =j_{x}j_{y},
\end{align}
where we defined the dimensionless couplings $j_{a}\propto\rho J_{a}$,
with $\rho=1/(\hbar v)$ representing the effective density of states
for the 1D bosonic environment.

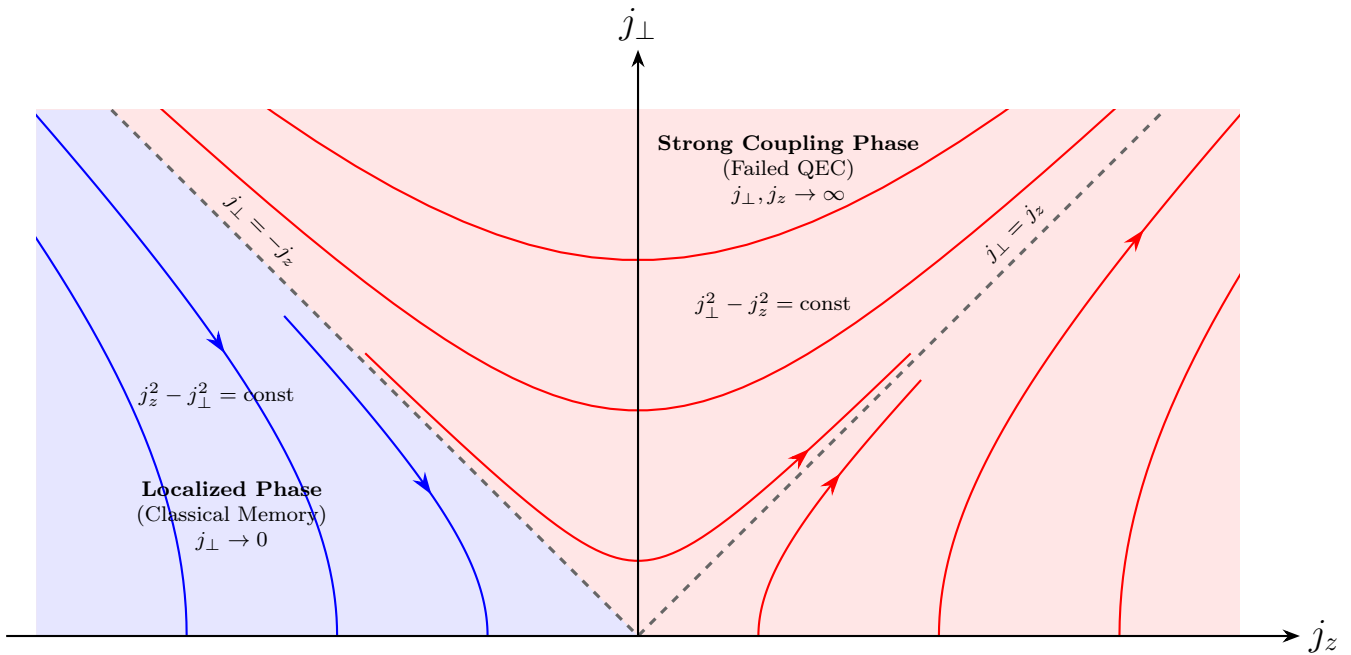
\begin{figure*}[t]
	\centering
	\tikzset{
		flowarrow/.style={
			decoration={
				markings,
				mark=at position #1 with {\arrow{Stealth[length=3mm, width=2mm]}}
			},
			postaction={decorate}
		}
	}
	
	\begin{tikzpicture}[scale=2.0, >=Stealth, font=\footnotesize]
		\def\xmax{4}
		\def\ymax{3.5}
		
		\begin{scope}
			\clip (-\xmax, 0) rectangle (\xmax, \ymax);
			
			\fill[red!10] (-\xmax, 0) rectangle (\xmax, \ymax);
			\fill[blue!10] (0,0) -- (-\ymax, \ymax) -- (-\xmax, \ymax) -- (-\xmax, 0) -- cycle;
			
			\draw[dashed, very thick, black!60] (-\ymax, \ymax) -- (0,0) -- (\ymax, \ymax);
			
			
			\foreach \c in {1, 2, 3} {
				\draw[blue, thick, flowarrow=0.6] 
				plot[domain=1.5:0, samples=50] ({-1*\c*cosh(\x)}, {\c*sinh(\x)});
			}
			
			\foreach \k in {0.5, 1.5, 2.5} {
				\draw[red, thick, flowarrow=0.8] 
				plot[domain=-2:2, samples=50] ({\k*sinh(\x)}, {\k*cosh(\x)});
			}
			
			\foreach \c in {0.8, 2, 3.2} {
				\draw[red, thick, flowarrow=0.6] 
				plot[domain=0:1.5, samples=50] ({\c*cosh(\x)}, {\c*sinh(\x)});
			}
		\end{scope}
		
		
		\draw[->, thick] (-\xmax-0.2, 0) -- (\xmax+0.4, 0) node[right, font=\Large] {$j_z$};
		\draw[->, thick] (0, 0) -- (0, \ymax+0.4) node[above, font=\Large] {$j_\perp$};
		
		\node[black, rotate=45] at (2.5, 2.7) {$j_\perp = j_z$};
		\node[black, rotate=-45] at (-2.5, 2.7) {$j_\perp = -j_z$};
		
		\node[align=center, black] at (-2.7, 0.8) {
			\textbf{Localized Phase}\\
			(Classical Memory)\\
			$j_\perp \to 0$
		};
		
		\node[align=center, black] at (1.0, 3.1) {
			\textbf{Strong Coupling Phase}\\
			(Failed QEC)\\
			$j_\perp, j_z \to \infty$
		};
		
		\node[black] at (-2.8, 1.6) {$j_z^2 - j_\perp^2 = \text{const}$};
		\node[black] at (0.9, 2.2) {$j_\perp^2 - j_z^2 = \text{const}$};
		
	\end{tikzpicture}
	\caption{Kosterlitz-Thouless (KT) renormalization group flow}
	\label{fig:KT-flow}
\end{figure*}

These flow equations possess exact constants of motion, such as $j_{x}^{2}-j_{y}^{2}=\text{const}$
and $j_{z}^{2}-j_{x}^{2}=\text{const}$. In a symmetric error model
where the bare transverse errors are equal ($j_{x}=j_{y}\equiv j_{\perp}$),
the system simplifies to the iconic Kosterlitz-Thouless (KT) flow,
\begin{align}
\frac{\partial j_{\perp}}{\partial l} & =j_{\perp}j_{z},\\
\frac{\partial j_{z}}{\partial l} & =j_{\perp}^{2},
\end{align}
that is summarized in Fig.~\ref{fig:KT-flow}. 

While the derived theoretical model naturally supports anisotropic
couplings, for simplicity, we analyze a model where the initial couplings
are isotropic in magnitude

\begin{equation}
\left|J\right|=\lambda\sqrt{\frac{2L}{\pi}}\left(\bar{\lambda}^{2}\right)^{\frac{L}{4}}.
\end{equation}
Depending on the sign of the longitudinal coupling, there are two
distinct thermodynamic regions to consider. 

\subsection{The ferromagnetic phase at zero temperature}

The first, and less destructive, is the ferromagnetic regime ($J_{z}\leq-J_{\perp}$).
In this region, the RG flow drives the transverse coupling to zero
($J_{\perp}\to0$). While the longitudinal coupling $J_{z}$ undergoes
a small finite renormalization (or marginally flows to zero exactly
on the separatrix $J_{z}=-J_{\perp}$), the transverse tunneling events
are entirely suppressed. At long times, the system effectively reduces
to a pure dephasing model. Hence, the off-diagonal coherences of the
logical qubit are not exponentially destroyed, but instead decay asymptotically
as a power law

\begin{equation}
\rho_{\bar{\uparrow}\bar{\downarrow}}\left(t\right)\propto\left(\frac{\tau_{\text{QEC}}}{t}\right)^{2\left(j_{z}^{*}\right)^{2}},
\end{equation}
where $j_{z}^{*}$ is the renormalized fixed-point coupling. Because
the quantum information is not actively scrambled by bit-flips, the
logical qubit in this phase effectively degrades into a classical
memory at $t\to\infty$.

While a power-law decay fundamentally lacks a characteristic exponential
lifetime, such as the standard $T_{2}$ time found in Markovian open
quantum systems, we can still establish an operational timescale for
the QEC protocol by defining a strict logical error threshold. Let
us assume that the quantum algorithm demands the accumulated phase
error probability remains below a critical value $\epsilon$ (for
instance, a $1\%$ loss of coherence, $\epsilon=0.01$). By setting
the survival probability $\rho(t_{\text{mem}})/\rho(0)=1-\epsilon$,
we can invert the asymptotic decay to extract the effective lifetime
of the logical memory

\begin{equation}
t_{\text{mem}}=\tau_{\text{QEC}}(1-\epsilon)^{-\frac{1}{2(j_{z}^{*})^{2}}}\approx\tau_{\text{QEC}}\exp\left(\frac{\epsilon}{2(j_{z}^{*})^{2}}\right),
\end{equation}
where the approximation holds for strict error thresholds ($\epsilon\ll1$).
This expression beautifully illustrates the physical mechanism of
QEC protection in the ferromagnetic regime. Even though the temporal
decay is algebraic rather than exponential, the active error correction
drives the renormalized coupling $j_{z}^{*}$ to an exponentially
small value. Therefore, the operational lifetime of the quantum memory
becomes exponentially prolonged relative to the fundamental clock
speed $\tau_{\text{QEC}}$ of the classical decoder.

\subsection{The antiferromagnetic phase at zero temperature}

The second, fundamentally destructive, case is the antiferromagnetic
regime ($J_{z}>-J_{\perp}$), which includes the strictly isotropic
limit $J_{z}=J_{\perp}>0$. In this region, the renormalization group
flow inevitably runs away to strong coupling. Even if the initial
QEC error rates are highly anisotropic, the flow trajectories universally
converge, dynamically restoring an $SU(2)$ symmetry as the system
approaches the infrared fixed point. Physically, this strong-coupling
fixed point is characterized by the formation of a Kondo singlet,
that corresponds to the maximal entanglement between the logical qubit
and the environment. 

In the context of the surface code, this means the logical Hilbert
space becomes maximally entangled with the continuous bosonic environment,
completely scrambling the stored quantum information. The characteristic
energy scale at which this macroscopic entanglement occurs is the
celebrated Kondo temperature, $T_{K}$. For the isotropic coupling,
integrating the RG equations yields this non-perturbative energy scale
\begin{equation}
k_{B}T_{K}\approx\frac{\hbar}{\tau_{\text{QEC}}}\exp\left(-\frac{1}{j}\right),
\end{equation}
where $j=\rho J$ is the initial dimensionless coupling strength,
and the inverse QEC cycle time $\hbar/\tau_{\text{QEC}}$ serves as
the fundamental ultraviolet energy cutoff of the bath.

By Wick rotating the partition function from imaginary time back to
real time, we can define the characteristic timescale for the formation
of this singlet, and thus the definitive failure time of the quantum
memory, as

\begin{equation}
t_{K}\approx\frac{\hbar}{k_{B}T_{K}}\approx\tau_{\text{QEC}}\exp\left(\frac{1}{j}\right).
\end{equation}
This timescale dictates the absolute operational limit of the actively
corrected logical qubit in the strong-coupling regime. Unlike the
infinitely long, algebraic decay found in the ferromagnetic phase,
crossing the boundary into the antiferromagnetic phase guarantees
the inevitable, non-perturbative destruction of the logical quantum
state within a finite characteristic time $t_{K}$.

To explicitly determine the operational limits of the actively corrected
quantum memory, we substitute the exact macroscopic expression for
the initial coupling $J$ into the dimensionless parameter $j=\rho J=J/(\hbar v)$

\begin{equation}
j(L)=\frac{\lambda}{\hbar v}\sqrt{\frac{2L}{\pi}}\left(\bar{\lambda}^{2}\right)^{\frac{L}{4}}.
\end{equation}

The absolute Kondo timescale $t_{K}$ represents the catastrophic
entanglement of the logical qubit with the environment. However, for
most practical purposes, just as in the ferromagnetic case, this ultimate
theoretical boundary is not the relevant practical metric for executing
a quantum algorithm.

Quantum algorithms require the accumulated logical error probability
to remain strictly below a defined operational threshold $\epsilon\ll1$.
Long before the system forms a maximally entangled Kondo singlet at
$t\sim t_{K}$, the continuous environment perturbatively degrades
the logical fidelity. By approximating the early-time coherence decay
governed by the macroscopic failure rate $\Gamma_{L}\sim1/t_{K}$,
we can set the survival probability to $1-\epsilon\approx\exp(-t_{\text{comp}}/t_{K})$.
Inverting this relation for a strict algorithmic tolerance extracts
the practical computational time,
\begin{equation}
t_{\text{comp}}\approx\epsilon\,t_{K}\approx\epsilon\,\tau_{\text{QEC}}\exp\left(\frac{1}{j(L)}\right).
\end{equation}
This expression grounds the abstract formation of the Kondo singlet
in practical quantum engineering. It demonstrates that while $t_{K}$
marks the physical death of the logical qubit, the operational algorithmic
window $t_{\text{comp}}$ is a heavily truncated fraction dictated
by the required precision $\epsilon$. 

To evaluate the possible regimes of an actively corrected quantum memory,
we must substitute the dimensionless coupling $j(L)$ into the expression
for the Kondo failure time, $t_{K}\approx\tau_{\text{QEC}}\exp(1/j)$.
The asymptotic survival of the quantum information as the lattice
size $L\to\infty$ depends strictly on whether the effective coupling
$j(L)$ flows to zero or diverges. Analyzing the three distinct spatial
scaling regimes of the environmental correlation function reveals
three fundamentally different behaviors.

\paragraph{The Short-Range Regime ($z>1/2$): The Thermodynamic Threshold}

When the spatial correlations of the bath decay sufficiently fast
($z>1/2$), the base of the exponential scaling in $j(L)$ is a size-independent
constant, $\bar{\lambda}^{2}\propto16\lambda^{2}\tau_{\text{QEC}}^{2}/(\hbar^{2}a_{0}^{2(1-z)}a^{2z})$.
In this regime, the system exhibits a true thermodynamic threshold.
If the intrinsic environmental coupling is sufficiently weak such
that $\bar{\lambda}^{2}<1$, the bare Kondo coupling, $J(L)$, is
driven to the unstable $J=0$ fixed point of the RG equations as we
take $L\to\infty$.

For a finite $L$, the operation time of the memory is a double exponential
in $L$, 

\[
t_{\text{comp}}\approx\epsilon\,\tau_{\text{QEC}}\exp\left[\frac{\hbar v}{\lambda}\sqrt{\frac{\pi}{2L}}\left(\frac{\hbar a_{0}^{(1-z)}a^{z}}{4\lambda\tau_{\text{QEC}}}\right)^{\frac{L}{2}}\right],
\]
that will crucially depend on the velocity of the environment (that
in the traditional Kondo problem would be related to the density of
states of the fermionic environment at the Fermi level).

\paragraph{The Critical Regime ($z=1/2$)}

In the critical regime ($z=1/2$), the long-range spatial pairings
introduce a logarithmic divergence into the effective coupling base,
scaling as $\bar{\lambda}_{1/2}^{2}\propto\ln L$. Because the natural
logarithm grows monotonically with system size, there is no strict
asymptotic threshold. Even if the intrinsic environmental coupling
$\lambda$ is initially small, scaling up the code distance guarantees
that $\bar{\lambda}_{z=1/2}^{2}$ will eventually exceed unity. Once
this occurs, the bare Kondo coupling $J(L)$ is driven away from the
unstable $J=0$ fixed point and flows inevitably toward the strong-coupling
fixed point as $L\to\infty$. As a consequence, the operation time
of the memory asymptotically collapses to the fundamental clock speed
of the QEC cycle.

The logarithmic breach dictates that the surface code is not a true
topological phase of matter against $z=1/2$ noise; as the computer
scales toward the fault-tolerant limit, the required precision for
the physical error rate $\lambda$ must become increasingly stringent
to maintain the same level of logical protection. However, for any
reasonable $L$, the computational time becomes

\[
t_{\text{comp}}\approx\epsilon\,\tau_{\text{QEC}}\exp\left[\frac{\hbar v}{\lambda}\sqrt{\frac{\pi}{2L}}\left(\frac{\hbar}{4\lambda\tau_{\text{QEC}}}\sqrt{\frac{a_{0}a}{\ln L}}\right)^{\frac{L}{2}}\right],
\]
that is essentially the same as for the $z>1/2$ case. 

\paragraph{Long Range Regime ($z<1/2$)}

For highly non-local baths ($z<1/2$), the spatial correlations decay
so slowly that the base of the exponential scaling itself diverges
as a power law, $\bar{\lambda}_{z}^{2}\propto L^{1-2z}$, hence there
is also formally no threshold as $L\to\infty$. For a finite $L$,
the computational time is 

\[
t_{\text{comp}}\approx\epsilon\,\tau_{\text{QEC}}\exp\left[\frac{\hbar v}{\lambda}\sqrt{\frac{\pi}{2L}}\left(\frac{\hbar a_{0}^{(1-z)}a^{z}}{4\lambda\tau_{\text{QEC}}L^{1-2z}}\right)^{\frac{L}{2}}\right],
\]

\subsection{Superconducting Circuits}

State-of-the-art superconducting processors, such as fixed-lattice
transmons \citep{vezvaee_surface_2025}, currently lead the field
in raw operational speed. Driven by fast microwave pulses, the classical
QEC cycle is remarkably short, typically on the order of $\tau_{\text{QEC}}\approx1\text{ \ensuremath{\mu}s}$\citep{google_quantum_ai_suppressing_2023}.
However, the long-time coherence of these devices may fundamentally
be limited by environmental noise\citep{aumentado_quasiparticle_2023}.
While isolated superconducting qubits can be carefully tuned to \textquotedbl sweet
spots\textquotedbl{} to become first-order insensitive to low-frequency
$1/f$ flux and charge noise, scaling to a macroscopic surface code
fundamentally breaks this isolation\citep{paladino_1_2014,dutta_low-frequency_1981}.

In a lattice of hundreds or thousands of junctions\citep{google_quantum_ai_suppressing_2023},
inherent fabrication variations make it statistically impossible for
all qubits to reside simultaneously at their optimal tuning points.
Consequently, the macroscopic ensemble of off-sweet-spot qubits is
unavoidably subjected to an aggregate background of low-frequency
noise. If the constituent junctions are of excellent quality, the
collective distribution of their local fluctuators can effectively
change the noise power spectrum. Rather than the strict $1/f$ profile
characteristic of individual \textquotedbl dirty\textquotedbl{} junctions,
this vast ensemble averaging should drive the infrared limit toward
an Ohmic environment. Note that, spatially, the resonators could interact
with several junctions, hence this qubit crosstalk should be modeled
by a $z<1$, dynamical exponent. Specifically, capacitive crosstalk
across a two-dimensional transmon array fundamentally couples the
logical qubit to the charge density of the underlying substrate. This
mechanism was recently highlighted in studies of correlated charge
bursts induced by gamma-ray and muon collisions\citep{wilen_correlated_2021}.
While modern designs minimize direct crosstalk through increased physical
separation, such macroscopic footprints simultaneously enhance the
coupling to local charge traps, surface states, and piezoelectric
interactions with substrate phonons. This fact allows for a plausible
scenario for a strongly correlated long-wavelength environment. In
a 2D interacting fermionic plasma, the unscreened Coulomb interaction
yields gapless collective charge excitations (plasmons) with a dispersion
relation $\omega_{q}\propto\sqrt{q}$\citep{ando_electronic_1982}.
This square-root dispersion perfectly manifests a dynamical exponent
of $z=1/2$, plunging the macroscopic surface code directly into the
critical regime where the logarithmic divergence weaponizes the bath.
While definitive noise spectra for processors of this scale must ultimately
be established by future experiments, modeling this aggregate macroscopic
background as an Ohmic bath in the infrared limit is a physically
plausible foundation to explore its bounds.

To prevent parasitic capacitive crosstalk, transmons must be physically
large, requiring a macroscopic lattice pitch of roughly $a\approx1\text{ mm}$\citep{google_quantum_ai_suppressing_2023,vezvaee_surface_2025}.
According to our figure of merit, scaling the surface code in this
architecture creates a massive spatial footprint (a distance $L=30$
code spans several centimeters). We can directly evaluate the viability
of this architecture by applying these hardware parameters to our
critical threshold limits. If the macroscopic noise background remains
strictly short-range ($z>1/2$), the exceptionally fast clock speed
$\tau_{\text{QEC}}\approx1\text{ }\mu\text{s}$ dominates the denominator
of $\lambda_{c}$, establishing a highly forgiving and size-independent
error threshold. However, if unmitigated crosstalk pushes the environment
into the critical ($z=1/2$) or long-range ($z<1/2$) regimes, the
massive physical footprint weaponizes the continuous bath. As the
lattice scales, the scaling penalty $L^{1/2-z}$ in the denominator
will rapidly drive the critical coupling $\lambda_{c}$ toward zero,
forcing an impossibly stringent requirement on the bare junction quality
to avoid the non-perturbative Kondo collapse.

\subsection{Neutral Atom Arrays}

In stark contrast to superconducting circuits, reconfigurable neutral
atom arrays present a fundamentally different noise profile\citep{saffman_quantum_2010,saffman_quantum_2016,morgado_quantum_2021,bluvstein_logical_2024}.
In these architectures, the continuous execution of the surface code
requires physically shuttling atoms between entangling and readout
zones using optical tweezers\citep{bluvstein_logical_2024}. This
mechanical overhead limits the QEC cycle time to significantly slower
bounds, typically on the order of $\tau_{\text{QEC}}\approx1\text{ ms}$\citep{bluvstein_logical_2024}.

While the macroscopic phase noise of the control lasers constitutes
the primary source of correlated errors across the atomic array, this
driving field is explicitly classical. It subjects the atoms to stochastic
dephasing, a process perfectly captured by the standard, discrete
theory of topological quantum error correction. Because classical
threshold theorems were explicitly designed to suppress this exact
type of stochastic error propagation, classical laser noise does not
trigger the macroscopic infrared divergences central to our model.
Instead, the only true macroscopic quantum continuum with which the
neutral atoms interact is the vacuum electromagnetic field, which
now becomes the strict focus of our scaling limits.

Crucially, strong coupling to the continuous environment only occurs
during the brief, microsecond-long Rydberg excitations required to
perform two-qubit entangling gates. During this fleeting window, the
atoms become highly polarizable and interact with both the vacuum
electromagnetic continuum (via spontaneous emission) and ambient thermal
blackbody radiation. However, because the atoms spend the vast majority
of the millisecond QEC cycle in their dipole-isolated ground states,
their time-averaged effective microscopic coupling is phenomenologically
suppressed by orders of magnitude.

To precisely quantify this physical protection, we can evaluate the
thermodynamic threshold specifically for the $z=1$ electromagnetic
vacuum. Substituting $z=1$ into the short-range scaling base mathematically
eliminates the short-distance cutoff $a_{0}$, isolating the lattice
spacing $a$ as the sole ultraviolet spatial regulator
\[
\bar{\lambda}_{1}^{2}=16\left(\frac{\lambda\tau_{\text{QEC}}}{\hbar a}\right)^{2}.
\]
By defining the microscopic coupling relative to the speed of light,
$\lambda=g\hbar c$, the physics naturally reorganizes into a dimensionless
kinematic fraction, $\bar{\lambda}_{1}^{2}=16g^{2}(c\tau_{\text{QEC}}/a)^{2}$.
For a state-of-the-art array with a microscopic pitch of 3~$\mu\text{m}$
and a slow cycle time of 1 ms, the vacuum light cone traverses an
astonishing $10^{11}$ lattice sites during a single error correction
cycle. Setting the strict threshold constraint $\bar{\lambda}_{1}^{2}<1$
reveals that the array must maintain a time-averaged dimensionless
coupling of $g_{c}\approx2.5\times10^{-12}$. If the atoms were continuously
coupled to the environment, the massive phase space accumulated by
this slow macroscopic clock would instantly destroy the logical qubit.
Therefore, the survival of the neutral atom architecture relies entirely
on its fundamental isolation: by parking the atoms in strictly dipole-forbidden
ground states for the vast majority of the cycle, the hardware effortlessly
suppresses the time-averaged coupling well below this stringent $10^{-12}$
threshold.

It is crucial to emphasize that this derived coupling tolerance represents
a strict theoretical ceiling rather than a practical engineering target.
Our effective field theory isolates the thermodynamic threat of the
macroscopic bath by explicitly assuming that the classical QEC operations
(syndrome extraction, spatial shuttling, and decoding) are executed
instantaneously. In physical atomic and trapped-ion architectures,
these interventions are inherently slow, finite-time processes. During
the prolonged durations required to execute physical shuttling, laser-driven
entangling gates, and fluorescence readout, the internal qubit states
are heavily hybridized with their continuous environment, exposing
them to spontaneous emission, macroscopic laser phase noise, and motional
heating. Because our model assumes the system is perfectly projected
at each discrete clock tick and only vulnerable during the interstitial
free-evolution periods, it does not capture the active accumulation
of decoherence during these slow operational windows. Consequently,
the catastrophic Kondo limits derived here establish an absolute best-case
boundary: they mathematically guarantee that even if all discrete
logical interventions could be executed flawlessly, the macroscopic
combinatorics of the continuous vacuum will still violently destroy
the quantum memory if the time-averaged coupling exceeds these stringent
thresholds. Therefore, in the case of Neutral Atom Arrays for quantum
memory, the most relevant time for decoherence occurs within the QEC
procedure itself, which is not captured in our calculation.

\section{Finite Temperature Results}

Up to this point, our analytical derivation has treated the continuous
macroscopic environment as strictly residing at zero temperature.
We emphasize that this zero-temperature calculation serves as a necessary
theoretical baseline to establish the interacting infrared field theory
and the geometric threshold $L_{c}$. However, the operational reality
of active topological quantum error correction inherently precludes
this idealized ground state. The continuous execution of the QEC cycle,
relying on the physical application of quantum gates, the intrinsically
dissipative process of syndrome measurements, and the subsequent classical
feedback, constantly pumps energy into the lattice of physical qubits.
Because these physical qubits are inextricably coupled to the underlying
continuous phase field, every localized operation unavoidably injects
bosonic excitations into the surrounding macroscopic environment.

From a spectral perspective, creating highly localized pulses in time,
such as ultrafast gates and projective measurements, inherently requires
a broad superposition of frequencies. Consequently, every discrete
QEC operation unavoidably excites a continuous spectrum of bath modes,
directly injecting energy into the low-frequency infrared sector.
While the vast majority of this energy resides in the high-frequency
ultraviolet modes, which are more easily thermalized and extracted
by the physical cooling apparatus, the low-frequency modes are continuously
pumped by the ongoing algorithmic cycle. The accumulation of these
low-frequency excitations actively scrambles the continuous environment.
In our framework, the assumption of a constant effective temperature
$T$ models the steady-state thermodynamic equilibrium between this
inevitable measurement-induced heating and the finite cooling power
of the macroscopic refrigerator. Nevertheless, nothing in the fundamental
physics precludes a more complex, non-equilibrium scenario where the
relentless operation of the classical decoder causes the macroscopic
environment to progressively heat up over the computational runtime.

The infrared model that we derived, Eq.~(\ref{eq:logical_kondo-1}),
is a boundary conformal field theory\citep{affleck_kondo_1991,di_francesco_conformal_1997}.
Hence, by employing the standard conformal mapping from the infinite
complex plane to a cylinder of circumference $\beta=\hbar/(k_{B}T)$,
we can rigorously promote the zero-temperature correlation functions
to finite temperature. For the spatial gradient of the phase field,
which acts as a primary field with scaling dimension $\Delta=1$,
the real-time two-point correlator becomes

\[
\langle\partial_{x}\hat{\phi}(0,t)\partial_{x}\hat{\phi}(0,0)\rangle_{T}\propto\left(\frac{\pi k_{B}T/\hbar}{\sinh\left(\frac{\pi k_{B}T}{\hbar}t\right)}\right)^{2}.
\]
Similar expressions hold for the vertex operators $\exp(\pm i\sqrt{4\pi}\theta(0,0))$
corresponding to the transverse logical errors, as the fundamental
duality of the continuous QEC model dictates that they share the exact
same conformal dimension.

The finite temperature $T$ introduces a fundamental thermal energy
scale into the problem, acting as an absolute infrared cutoff that
halts the renormalization group flow. In the zero-temperature limit,
the RG trajectories proceed indefinitely toward $t\to\infty$. However,
at finite temperature, the macroscopic correlation length is strictly
bounded by the thermal wavelength, and the RG flow is effectively
terminated at the characteristic thermal timescale $t_{\text{th}}=\hbar/(\pi k_{B}T)$.
This thermal cutoff drastically modifies the asymptotic behavior of
the logical qubit in both regimes.

\subsection{The ferromagnetic phase at finite temperature}

In the topologically protected ferromagnetic regime ($J_{z}\leq-J_{\perp}$),
the zero-temperature analysis revealed that the off-diagonal coherences
of the logical qubit decay asymptotically as a power law, yielding
a practically infinite $T_{2}$ lifetime. However, by substituting
the finite-temperature conformal correlators into the evaluation of
the logical coherence, the asymptotic behavior is fundamentally transformed.
For times exceeding the thermal scale ($t\gg t_{\text{th}}$), the
hyperbolic sine function transitions into a purely exponential decay.
The previously algebraic survival probability becomes bounded by a
finite thermal relaxation rate,
\[
\rho_{\bar{\uparrow}\bar{\downarrow}}(t)\propto\exp\left(-2\pi\left(j_{z}^{*}\right){}^{2}\frac{k_{B}T}{\hbar}t\right).
\]
This result implies that the idealized \textquotedbl classical memory\textquotedbl{}
of the zero-temperature fixed point is destroyed by the active heating
of the QEC protocol. The residual fixed-point coupling $j_{z}^{*}$,
which was benign at $T=0$, now mediates a steady absorption and emission
of thermal bath excitations, leading to a finite exponential lifetime $T_{2}^{\text{th}}\propto\hbar/[k_{B}T(j_{z}^{*})^{2}]$.
While the QEC protocol still exponentially suppresses the effective
coupling $j_{z}^{*}$ through the geometric scaling $L$, the introduction
of a finite temperature strictly re-establishes Markovian-like exponential
decoherence at long times for any finite $L$.

\subsection{The antiferromagnetic phase at finite temperature}

Conversely, in the antiferromagnetic regime ($J_{z}>-J_{\perp}$),
the finite temperature introduces a profound physical competition
between the Kondo energy scale $k_{B}T_{K}$ and the thermal energy
$k_{B}T$. The fate of the quantum memory is now determined by which
energy scale is reached first during the RG flow.

If the environment remains relatively cold ($T\ll T_{K}$), the RG
flow reaches the strong-coupling fixed point before the thermal cutoff
intervenes. The Kondo singlet successfully forms, and the logical
qubit can suffer the same catastrophic, non-perturbative entanglement
collapse derived in the zero-temperature case.

However, if the relentless execution of the QEC gates heats the macroscopic
environment such that $T\gg T_{K}$, the thermal fluctuations violently
interrupt the formation of the Kondo screening cloud. The RG flow
is halted in the perturbative weak-coupling regime. Although, the thermal
interruption \textquotedbl saves\textquotedbl{} the logical qubit from the maximally entangled Kondo collapse, the quantum
information is nonetheless destroyed
by standard thermal relaxation. The bath excitations scramble the
logical state via Korringa-like relaxation processes. The resulting
exponential decay rate $\Gamma_{K}$ is strictly linear in temperature
and scales with the square of the macroscopic dimensionless coupling,
\[
\Gamma_{K}\propto j(L)^{2}\frac{k_{B}T}{\hbar}.
\]
Depending on the value of $z$, the Korringa relaxation rate $\Gamma_{K}$
can diverge as the code distance $L$. Once again, this assures that
there is no thermodynamic threshold for these cases.

\section{Summary and Conclusions}

Traditional QEC relies heavily on large-scale classical simulation
because it models decoherence via discrete Pauli channels, which can
be efficiently simulated in polynomial time using the stabilizer formalism
(the Gottesman-Knill theorem). In stark contrast, our manuscript evaluates
the surface code coupled to a continuous, highly entangled bosonic
field theory. The combined Hilbert space of a macroscopic topological
lattice interacting with an infinite-dimensional gapless continuum
is classically intractable. Simulating the exact real-time dynamics
of this joint system would essentially require a fault-tolerant quantum
computer.

Because brute-force finite-size numerics are fundamentally incapable
of capturing the macroscopic infrared divergences of a continuous
bath, we introduce a theoretical paradigm shift by utilizing the exact
analytical machinery of effective field theory and the Wilsonian Renormalization
Group (RG). By mapping the surface code onto a 1+1D boundary conformal
field theory, we demonstrate that the traditional assumption of perfectly
isolated logical qubits critically fails in the macroscopic limit.
From the perspective of open quantum systems, a scalable quantum computer
is undeniably a macroscopic object. As the physical footprint of the
code expands to achieve higher logical distances, its combinatorial
coupling to the continuous environmental bath can diverge. By assuming
long-time correlations with Ohmic characteristics, we defined the
criterion for the existence of a threshold for all environments with
a dynamical exponent $z>1/2$. For a general environment, $J\left(\omega\right)\propto\omega^{s}$,
this criterion changes to
\[
z>\frac{1}{s+1}.
\]

In the Ohmic case, $s=1$, we use the standard RG analysis to evaluate
the computational time for all values of $z$ for finite code distance
$L$. We continued the discussion by using the conformal finite temperature
propagator to evaluate the effects of an environment at finite temperature.
The algebraic decay characteristic of the protected zero-temperature
phase is replaced by an exponential Korringa-like thermal relaxation.
Thus, even in the most forgiving short-range environments, active
QEC cannot provide an infinitely lived memory for a finite code distance
$L$.

The worst possible scenario for an actively corrected surface code
is a sub-Ohmic ($s<1$) environment. In this regime, our fundamental
RG equation changes to 
\[
\frac{dj}{dl}=\left(1-s\right)j.
\]
This relevant perturbation drastically reduces the operational lifetime
from an exponential dependence on the inverse coupling to a strict
power law, 
\[
t_{\text{comp}}\approx\epsilon\,t_{K}=\epsilon\tau_{\text{QEC}}\left[\frac{1}{j(L)}\right]^{\frac{1}{1-s}}.
\]
If a thermodynamic threshold exists (the short-range case, $z>1/\left(s+1\right)$),
the topological protection is heavily downgraded from a double exponential
to a simple exponential scaling in $L$. However, in both the critical
($z=1/\left(s+1\right)$) and highly correlated ($z<1/(s+1$)) regimes,
topological protection is completely destroyed. Because the bare coupling
$j(L)$ diverges with system size, the computational time inevitably
collapses as the lattice scales, decaying logarithmically in the critical
case and polynomially in the highly correlated regime.

These macroscopic bounds highlight a critical divergence in hardware
viability. For solid-state architectures like superconducting circuits,
which possess massive spatial footprints, scaling to large code distances
severely risks triggering non-perturbative infrared collapses if the
aggregate background noise exhibits critical or long-range spatial
correlations. In these regimes, the sheer physical size of the lattice
weaponizes the continuous bath, easily overwhelming their ultrafast
clock speeds. Conversely, atomic array architectures circumvent this
thermodynamic threat not through spatial compactness, but through
extreme temporal isolation. Because neutral atoms interact strongly
with the vacuum continuum only during fleeting Rydberg excitations,
they spend the vast majority of the classical QEC cycle decoupled
in their ground states. This fundamental temporal isolation heavily
suppresses their time-averaged continuous coupling, allowing them
to remain below the stringent macroscopic threshold even as the physical
lattice scales. Since their strong decoherence occurs during the QEC
protocol itself, a proper evaluation of the threshold for neutral
atoms cannot be directly read from our work.

Ultimately, this work bridges the historically distinct domains of
quantum information theory, condensed matter, and dissipative quantum
mechanics. It rigorously demonstrates that decoherence cannot be circumvented
by algorithmic error correction alone. To realize truly scalable fault-tolerant
quantum computation, future architectural designs must not only optimize
discrete gate fidelities, but must also address the residual logical
decoherence that always exists in a QEC protocol.
\begin{acknowledgments}
We would like to thank Prof. Amir Caldeira for invaluable 
discussions on quantum information, field theory and 
dissipative quantum mechanics.
\end{acknowledgments}

\section{Statements and Declarations}

\begin{itemize}
\item \emph{Funding}: This work was partially 
supported by the S\~{a}o Paulo Research Foundation (FAPESP), 
Brazil, Process No. 2022/15453-0 and by the Ministry
of Education, Singapore, under its Research Centre of
Excellence award to the Institute for Functional Intelligent
Materials, National University of Singapore
(I-FIM, project No. EDUNC-33-18-279-V12).
AI tools (Gemini) were used to improve the clarity and style of the manuscript.
These tools were applied to refine the text in accordance
with established principles of scientific writing \citep{Gopen_Swan-90}.

\item \emph{Competing Interests}: The authors declare no competing interests.

\item \emph{Authors’ contribution}: The authors equally contribute to this work.

\item \emph{Ethics Declaration:}: Not applicable.

\end{itemize}

\appendix

\section{\label{sec:An-adversary-noise}An Adversarial Noise Environment for
Quantum Information Storage}

Before detailing the formal derivation of the macroscopic effective
model from a microscopic coupling, it is crucial to clearly define
the spectral boundary of our theory and its relationship to standard
fault-tolerant assumptions.

In standard quantum information theory, environmental decoherence
is typically characterized by independent $T_{1}$ relaxation and
$T_{2}$ dephasing times. These metrics are strictly derived under
the Born-Markov approximation, which fundamentally assumes that the
bath correlation time is infinitesimally short. In the spectral language
of our field theory, these independent, localized Markovian errors
correspond exclusively to the high-frequency ultraviolet (UV) sector
of the environment. Because the active QEC cycle operates at an ultrafast
clock speed ($\tau_{\text{QEC}}$), it acts as a highly effective
temporal filter. The classical decoder successfully identifies and
corrects these fast, uncorrelated UV fluctuations, effectively masking
them from the long-time dynamics of the logical subspace.

However, any physical continuum inherently possesses a low-frequency
infrared (IR) tail. These slow, macroscopic bosonic modes evolve on
timescales much longer than the QEC cycle ($t\gg\tau_{\text{QEC}}$),
thereby severely violating the Born-Markov approximation. Because
they do not manifest as discrete, independent Pauli errors that the
classical decoder can cleanly resolve within a single cycle, standard
discrete QEC protocols are fundamentally blind to this slow IR entanglement.

The 1D Tomonaga-Luttinger liquid and the subsequent Kondo collapse
described in the main text do not replace these local Markovian errors;
rather, they represent the universal, non-Markovian low-energy physics
that strictly survives the active QEC temporal filter. The microscopic
derivation below demonstrates exactly how this uncorrected IR sector
dynamically entangles with the macroscopic spatial footprint of the
code, flowing universally into the effective continuous field theory
of Eq.~(\ref{eq:logical_kondo-1}).

Deriving the exact microscopic noise power spectrum for a macroscopic
quantum memory composed of thousands of physical components is theoretically
intractable; such precise characterizations must ultimately be determined
by future experiments. Nevertheless, we can construct a rigorous,
adversarial quantum environment based on two fundamental physical
principles: 1) quantum information is never fundamentally destroyed,
but rather irreversibly scrambled from the logical qubit into the
environment via continuous unitary entanglement; and 2) to act as
a true thermodynamic bath, the environment must possess propagating
degrees of freedom capable of irreversibly carrying this information
away from the local quantum memory.

To satisfy the first principle, we assume the environment is composed
of a macroscopic ensemble of environmental qubits, $\vec{s}(\vec{k})$,
which vastly outnumber the physical qubits of the surface code. These
environmental degrees of freedom interact with the physical qubits
of the surface code, $\vec{\sigma}(\vec{x}_{j})$, via a general anisotropic
exchange interaction

\begin{equation}
H_{\text{int}}=\sum_{jk}\left[\lambda_{x,k}s^{x}\left(\vec{k}\right)\sigma_{j}^{x}+\lambda_{y,k}s^{y}\left(\vec{k}\right)\sigma_{j}^{y}+\lambda_{z,k}s^{z}\left(\vec{k}\right)\sigma_{j}^{z}\right].
\end{equation}
This spin-exchange formulation serves as a universal representation
for any quantum informational bath.

To satisfy the second principle, we must assume that instead of a
static lattice, these environmental qubits are actively propagating.
Hence, the abstract label $\vec{k}$ represents the continuous momentum
of these itinerant degrees of freedom. A natural and mathematically
rigorous way to formalize this propagating bath is to represent each
environmental qubit as the spin degree of freedom of an itinerant
fermion with momentum $\vec{k}$. By assuming these environmental
fermions do not interact with one another, the adversarial bath formally
maps to a non-interacting Fermi gas that is vastly larger than the
surface code.

This adversarial environment allows us to directly derive the effective
IR noise model, bypassing the derivation in the main text. For the
sake of completeness, we also provide a formal integration of the
high-frequency modes in subsection~(\ref{subsec:The-Ultraviolet-Model}).

\subsection{The Infrared Model}

To proceed with the exact mathematical mapping, we follow the standard
field-theoretic treatment of magnetic impurities in metals. It is
crucial here to strictly define the order of thermodynamic limits.
In the theory of open quantum systems, the thermodynamic limit of
the continuous bath ($V_{\text{bath}}\to\infty$) must be taken strictly
prior to scaling the macroscopic footprint of the logical qubit ($L\to\infty$).
Because the bath volume is fundamentally infinite relative to the
system, the characteristic size of the surface code $L$ remains negligible
compared to the infinitely long wavelengths of the deep infrared (IR)
fermions ($k\to0$).

Because these relevant IR modes always satisfy the scattering condition
$kL\ll1$, the entire spatial footprint of the logical qubit acts
effectively as a localized scattering potential. Consequently, the
logical qubit couples predominantly to the spherically symmetric $s$-wave
scattering channel of the environment. By integrating out the higher-angular-momentum
modes that decouple from this localized interaction, a 2D or 3D Fermi
gas rigorously reduces to an effective 1D chiral fermion channel interacting
with the logical impurity at the origin\citep{hewson_kondo_2009}.

We can therefore express the environmental spin operators at the location
of the logical qubit using the 1D fermionic creation and annihilation
operators, $\psi_{s}^{\dagger}(0)$ and $\psi_{s}(0)$, where $s\in\{\uparrow,\downarrow\}$
denotes the spin projection

\begin{align}
s^{z}(0) & =\frac{1}{2}\left[\psi_{\uparrow}^{\dagger}(0)\psi_{\uparrow}(0)-\psi_{\downarrow}^{\dagger}(0)\psi_{\downarrow}(0)\right],\label{eq:sz_fermion}\\
s^{+}(0) & =\psi_{\uparrow}^{\dagger}(0)\psi_{\downarrow}(0),\label{eq:splus_fermion}\\
s^{-}(0) & =\psi_{\downarrow}^{\dagger}(0)\psi_{\uparrow}(0).\label{eq:sminus_fermion}
\end{align}

To extract the macroscopic IR dynamics of this fermionic bath, we
apply the standard identities of 1D Abelian bosonization\citep{von_delft_bosonization_1998}.
The fermionic fields can be expressed as coherent exponentials of
free bosonic phase fields

\begin{equation}
\psi_{s}(x)=\frac{1}{\sqrt{2\pi a}}\exp\left[-i\sqrt{\pi}\left(\Phi_{c}(x)+s\Phi_{s}(x)\right)\right],
\end{equation}
where $a$ is the short-distance cutoff, while $\Phi_{c}(x)$ and
$\Phi_{s}(x)$ are the independent bosonic fields corresponding to
the charge and spin density fluctuations of the environment, respectively.

A profound physical consequence of this 1D geometric reduction is
spin-charge separation. Because the logical qubit interacts exclusively
with the spin degrees of freedom of the environmental fermions (as
defined above), the charge field $\Phi_{c}(x)$ completely decouples
from the interaction Hamiltonian and can be trivially traced out.
The quantum dissipative dynamics are therefore entirely governed by
the environmental spin field, $\Phi_{s}(x)$, which perfectly maps
to the macroscopic Tomonaga-Luttinger phase field defined in our main
text. Crucially, the bosonic density-density correlation functions
of this bosonized Fermi gas inherently produce an Ohmic spectral density
($s=1$), perfectly matching the continuous macroscopic propagator
derived in Eq.~(\ref{eq:density-density}) of the main text.

Substituting the bosonized fermion operators into the environmental
spin densities evaluated at the impurity ($x=0$), we obtain the exact
bosonic representation of the bath

\begin{align}
s^{z}(0) & \propto\partial_{x}\phi(0),\\
s^{+}(0) & \propto\exp\left(i\sqrt{4\pi}\theta(0)\right),\\
s^{-}(0) & \propto\exp\left(-i\sqrt{4\pi}\theta(0)\right),
\end{align}
where $\theta(x)$ is the canonical conjugate momentum field to $\phi(x)$,
satisfying the standard commutation relation $[\phi(x),\partial_{y}\theta(y)]=i\delta(x-y)$.

By recombining the raising and lowering operators into the Cartesian
transverse spin components, $s^{x}=\frac{1}{2}(s^{+}+s^{-})$ and
$s^{y}=\frac{1}{2i}(s^{+}-s^{-})$, we directly recover the continuous
vertex operators:

\begin{align}
s^{x}(0) & \propto\cos\left(\sqrt{4\pi}\theta(0)\right),\\
s^{y}(0) & \propto\sin\left(\sqrt{4\pi}\theta(0)\right).
\end{align}

In this deep IR description, the logical operators manifest as localized
scattering degrees of freedom that couple directly to the continuous
bosonic environment, hence yielding precisely the boundary conformal
field theory of Eq.~(\ref{eq:logical_kondo-1}).

This rigorous microscopic path establishes a fundamental theorem for
quantum error correction in continuous environments: regardless of
the specific microscopic details of the itinerant adversarial bath,
the macroscopic IR entanglement between the logical qubit and the
continuous environmental spin density universally converges to the
anisotropic Kondo model.

\subsection{The Ultraviolet Model\label{subsec:The-Ultraviolet-Model}}

Following the exact same reasoning from the previous section, we can
now focus on the microscopic lattice of physical qubits interacting
with the quantum adversarial environment. Because the surface code
consists of a spatial array of localized spins coupled to the adversarial
itinerant fermionic bath, the full microscopic Hamiltonian is formally
equivalent to a Kondo lattice model.

While the Kondo lattice possesses a rich thermodynamic phase diagram\citep{coleman_heavy_2015},
the simple fact that we assume to observe a quantum memory at the
end of the QEC cycle, $\tau_{\text{QEC}}$, tells us that we are in
the perturbative regime of the Kondo lattice.

In this perturbative regime, we can formally integrate out the fast
environmental modes with wavelengths on the order of the physical
qubit separation, $d$. Integrating out these itinerant fermions generates
an effective Ruderman-Kittel-Kasuya-Yosida (RKKY)\citep{ruderman_indirect_1954,kasuya_theory_1956,yosida_magnetic_1957}
exchange interaction between the physical qubits. For a non-interacting
Fermi gas, this induced interaction is governed by the non-local static
spin susceptibility of the bath, $\chi(\vec{r})$. In a 3D electron
gas, the RKKY interaction oscillates at the Fermi momentum ($2k_{F}$)
and decays with the characteristic spatial envelope $J_{\text{RKKY}}(d)\propto1/d^{3}$\citep{aristov_indirect_1997}.
Conversely, for a 2D electron gas, the spatial envelope scales as
$\propto1/d^{2}$\citep{aristov_indirect_1997}. 

Crucially, if we assume a 2D noise environment, such as an itinerant
electron gas confined to the physical plane of the surface code, the
$1/d^{2}$ spatial envelope of the RKKY interaction exactly matches
the equal-time spatial correlation profile of the Ohmic bosonic propagator
assumed in Eq.~(\ref{eq:density-density}) of the main text. Therefore,
the perturbative UV integration of the high-frequency modes naturally
yields the exact macroscopic intra-cycle correlations required by
the continuous theory, formally driving the combinatorially enhanced
infrared coupling $J_{z}$ derived in Eq.~(\ref{eq:J_z}).

It is critical to note that this Ohmic ($z=1$) equivalence relies
strictly on the physical qubits coupling to the specific choice of
the $k$-dependence of the coupling constants. By changing this dependence,
it is possible to generalize this adversarial environment to represent
a strict threat to the QEC threshold.

\end{document}